\documentclass[traditabstract]{aa}

\usepackage{amsmath}
\usepackage{amssymb}
\usepackage{txfonts}
\usepackage{epsfig,graphicx}
\usepackage[colorlinks=true, linkcolor=blue, citecolor=blue, urlcolor=blue]{hyperref}
\usepackage{natbib}
\bibpunct{(}{)}{;}{a}{}{,} 
\begin{document}

\title{FitSKIRT: genetic algorithms to automatically fit dusty
  galaxies with a Monte Carlo radiative transfer code}
\titlerunning{FitSKIRT}
 
\author{Gert De Geyter
\and Maarten Baes
\and Jacopo Fritz
\and Peter Camps
}

\institute{Sterrenkundig Observatorium, Universiteit Gent, Krijgslaan 281-S9, B-9000 Gent, Belgium}

\date{\today}

\abstract{We present FitSKIRT, a method to efficiently fit radiative  
transfer models to UV/optical images of dusty galaxies. { These images 
have the advantage that they have better spatial resolution compared 
to FIR/submm data.} 
FitSKIRT  uses the GAlib genetic algorithm library to optimize the output 
of  the SKIRT Monte Carlo radiative transfer code. Genetic algorithms  
prove to be a valuable tool in handling the  multi-
dimensional search space as well as the noise induced by the  random 
nature of the Monte Carlo radiative transfer code. FitSKIRT  is tested on 
artificial images of a simulated edge-on spiral galaxy,  where we
gradually increase the number of fitted parameters. We find  that we can 
recover all model parameters, even if  all 11 model parameters are left 
unconstrained. Finally, we apply  the FitSKIRT code to a V-band image 
of the edge-on spiral galaxy  NGC\,4013. This galaxy has been modeled 
previously by other authors  using different combinations of radiative 
transfer codes and  optimization methods. Given the different models 
and techniques and  the complexity and degeneracies in the parameter 
space, we find  reasonable agreement between the different models. We 
conclude that  the FitSKIRT method allows comparison between different 
models and  geometries in a quantitative manner and minimizes the 
need of human  intervention and biasing. The high level of automation 
makes it an  ideal tool to use on larger sets of observed data.}


\maketitle

\section{Introduction}

In the past few years, the importance of the interstellar dust medium
in galaxies has been widely recognized. Dust regulates the physics and
chemistry of the interstellar medium, and reprocesses about one third
of all stellar emission to infrared and submm emission. Nevertheless, the
amount, spatial distribution and physical properties of the dust
grains in galaxies are hard to nail down.

The most direct method to trace the dust grains in galaxies is to
measure the thermal emission of the dust grains at mid-infrared (MIR),
far-infrared (FIR) and submm wavelengths. Obtaining total dust masses
from MIR, FIR or submm images should in principle be straightforward,
as dust emission is typically optically thin at these wavelengths and
thus total dust masses can directly be estimated from the observed
spectral energy distribution. There are a number of problems with this
approach, however. One complication is the notoriously uncertain value
of the dust emissivity at long wavelengths \citep{2003A&A...399L..43B,
  2003A&A...399.1073K, 2004A&A...425..109A, 2011ApJ...728..143S} and
the mysterious anti-correlation between the dust emissivity index and
dust temperature \citep{2003A&A...404L..11D, 2009ApJ...696..676S,
  2010ApJ...713..959V, 2012A&A...539A..71J, 2012A&A...541A..33J,
  2012ApJ...752...55K, 2012ApJ...756...40S}. A second problem is the
difficulty to observe in the FIR/submm window, which necessarily needs
to be done from space using cryogenically cooled instruments. Until
recently, the available FIR instrumentation was characterized by
limited sensitivity and wavelength coverage, and the submm window was
largely unexplored. This has now partly changed thanks to the launch
of the {\it Herschel} \citep{2010A&A...518L...1P} satellite, but also
this mission has a very limited lifetime. Finally, the third and most
crucial limitation is the poor spatial resolution of the available
FIR/submm instruments (typically tens of arcsec). This drawback is
particularly important if we want to determine the detailed
distribution of the dust in galaxies rather than just total dust
masses. This poor spatial resolution limits a detailed study of the
dust medium to the most nearby galaxies only
\citep[e.g.][]{2010A&A...518L..71M, 2010A&A...518L..65B,
  2012MNRAS.419.1833B, 2012A&A...543A..74X, 2012MNRAS.421.2917F,
  2012MNRAS.419..895D, 2012A&A...546A..34F, 2012ApJ...756...40S,
  2012MNRAS.425..763G}. Moreover, several authors have demonstrated
that even estimating total dust masses from global fluxes induces an
error due to resolution effects \citep{2011A&A...536A..88G,
  2012MNRAS.425..763G}.

The alternative method to determine the dust content in galaxies is to
use the extinction effects of the dust grains on the stellar emission
in the UV, optical or near-infrared (NIR) window. This wavelength
range has the obvious advantages that observations are very easy and
cheap, and that the spatial resolution is an order of magnitude better
than in the FIR/submm window. Furthermore, the optical properties of the dust
are much better determined in the optical than at FIR/submm
wavelengths. The main problem in using this approach is the difficulty
to translate attenuation measurements from broadband colors to actual
dust masses, mainly because of the complex and often counter-intuitive
effects of the star/dust geometry and multiple scattering
\citep{1989MNRAS.239..939D, 1992ApJ...393..611W, 1994ApJ...432..114B,
  1995MNRAS.277.1279D, 2001MNRAS.326..733B,
  2005MNRAS.359..171I}. Simple recipes that directly link an
attenuation to a dust mass are clearly not sufficient; the only way to
proceed is to perform detailed radiative transfer calculations that do
take into account the necessary physical ingredients (absorption,
multiple anisotropic scattering) and that can accommodate realistic
geometries. Fortunately, several powerful 3D radiative transfer codes
with these characteristics have recently been developed
\citep[e.g.][]{2001ApJ...551..269G, 2003MNRAS.343.1081B,
  2011ApJS..196...22B, 2006MNRAS.372....2J, 2008A&A...490..461B,
  2011A&A...536A..79R}.

A state-of-the-art radiative transfer code on itself, however, is not
sufficient to determine the dust content of galaxies from
UV/optical/NIR images. A radiative transfer simulation typically starts
from a 3D distribution of the stars and the dust in a galaxy model and
calculates how this particular system would look for an external
observer at an arbitrary viewing point, i.e.\ it simulates the
observations. The inverted problem, determining the 3D distribution of
stars and dust from a given reference frame, is a much harder nut to
crack. It requires the combination of a radiative transfer code and an
optimization procedure to constrain the input parameters. Several
fitting codes that combine a radiative transfer code with an
optimization algorithm have been set up \citep{1997A&A...325..135X,
  2005A&A...434..167S, 2007ApJS..169..328R, 2007A&A...471..765B,
  2012ApJ...746...70S}.  All too often, however, this optimization
procedure is neglected and {\em{chi-by-eye}} models are presented as
reasonable alternatives.

To optimize this given problem it is important to realize that the
parameter space is quite large, easily going up to 10 free parameters
or more. As discussed before, the complexity of the dust/star geometry
and scattering off dust particles is often counter-intuitive and
results in a non-linear, non-differentiable search space with multiple
local extrema . Furthermore, there is another inconvenient feature
induced by the radiative transfer code. Most state-of-the-art
radiative transfer codes are Monte Carlo codes where the images are
constructed by detecting a number of predefined photon
packages. Because of the intrinsic randomness of the Monte Carlo code,
images always contain a certain level of Poisson noise. If one runs a
forward Monte Carlo radiative simulation, it is usually
straightforward to make the number of photon packages in the
simulation so large that this noise becomes negligibly small. However,
if we want to couple the radiative transfer simulation to an
optimization routine, typically many thousands of individual
simulations need to be performed, which inhibits excessively long run
times for each simulation and hence implies that there will always be
some noise present. As a result, we have to fit noisy models to noisy
data, as reduced CCD data will always contain a certain level of noise
\citep{1991PASP..103..122N}.

For complex optimization problems like this, it is not recommended to
use classical optimization methods like the downhill simplex or
Levenberg-Marquardt methods, but rather apply a stochastic
optimization method instead. The main advantage of stochastic methods
such as simulated annealing, random search, neural networks, and
genetic algorithms over most classical optimization methods is their
ability to leave local extrema and search over broad parameter spaces
\citep{2003MNRAS.346..825F, 2001A&A...370..365T,
  2012arXiv1202.1643R}. Genetic algorithms \citep{Goldberg89,
  Mitchel98} are one class of the stochastic optimization methods that
stands out when it comes to noisy handling because it works on a set
of solutions rather than iteratively progressing from one point to
another. Genetic algorithms are often applied to optimize noisy
objective functions \citep{2000ApJ...545..974M,
  2003AJ....125.1958L}. During the past decade, genetic algorithms
have become increasingly more popular in numerous applications in
astronomy and astrophysics ranging from cosmology and gravitational
lens modeling to stellar structure and spectral fitting
\citep{1995ApJS..101..309C, 2000ApJ...545..974M, 2001A&A...370..365T,
  2003AJ....125.1958L, 2003MNRAS.346..825F, 2006MNRAS.367.1209L,
  2010A&A...516A..45B, 2012ApJ...746...70S, 2012MNRAS.427.1755R}. For
a recent overview of the use of genetic algorithms in astronomy and
astrophysics, see \citet{2012arXiv1202.1643R}.

In this paper, we present FitSKIRT, a tool designed to efficiently
model the stellar and dust distribution in dusty galaxies by fitting
radiative transfer models to UV/optical/NIR images. The optimization
routine behind the FitSKIRT code is based on genetic algorithms, which
implies that the code is able to efficiently explore large, complex
parameter spaces and conveniently handle the noise induced by the
radiative transfer code. Bias by human intervention is kept to an
absolute minimum and the results are determined in an objective
manner. In Section~{\ref{FitSKIRT.sec}} the radiative transfer code
and the genetic algorithm are discussed in more detail. We also
apply the genetic algorithms library on a complex but analytically
tractable optimization problem to check its reliability and
efficiency. This knowledge is then used to explain the major steps and
features of the FitSKIRT program. In Section~{\ref{TestImages.sec}} we
test the FitSKIRT program on reference images of which the input
values are exactly known. This should allow us to have a closer look
at the complexity of the problem and FitSKIRTs ability to obtain
reasonable solutions in an objective way. In
Section~{\ref{NGC4013.sec}} we apply FitSKIRT to determine the
intrinsic distribution of stars and dust in the galaxy
NGC\,4013, and we compare the resulting model to similar models
obtained by \citet{2007A&A...471..765B} and
\citet{1999A&A...344..868X}. Section~{\ref{Discussion.sec}} sums up.

\section{FitSKIRT}
\label{FitSKIRT.sec}

\subsection{The SKIRT radiative transfer code}

\begin{figure*}
  \centering
  \includegraphics[width=0.65\textwidth]{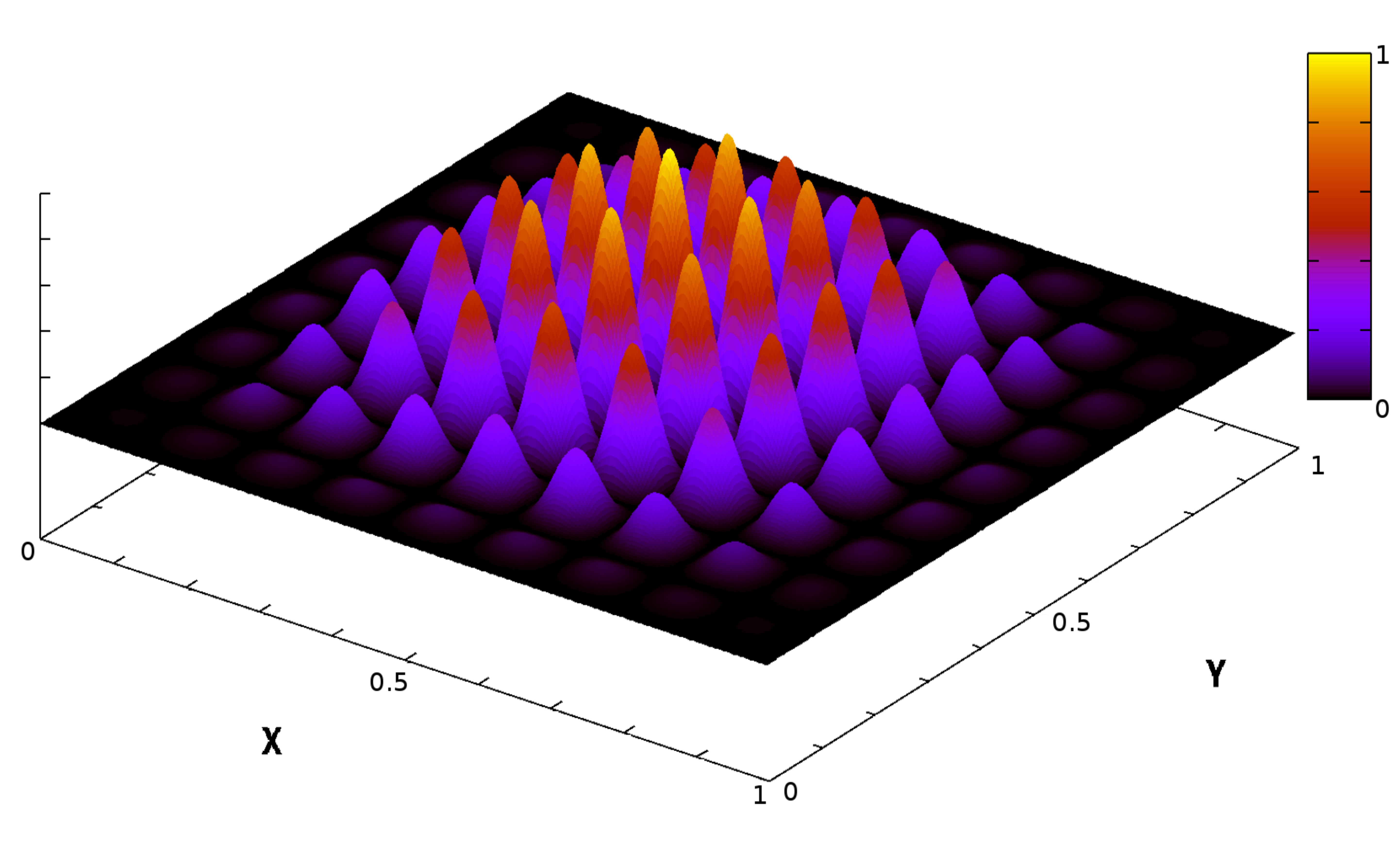} 
  \caption{The function defined by equation~(\ref{Charbonneau}) for
    $n=9$. It contains many steep local maxima surrounding the global
    maximum at $(x,y) = (0.5,0.5)$ and is therefore a strong test
    function for global optimization algorithms. }
  \label{Charbonneau.pdf}
\end{figure*} 

SKIRT \citep{2011ApJS..196...22B} is a 3D continuum Monte Carlo
radiative transfer code, initially developed to investigate the
effects of dust absorption and scattering on the observed gas and
stellar kinematics of dusty galaxies
\citep{2002MNRAS.335..441B,2003MNRAS.343.1081B}. The code has
continuously been adapted and upgraded to a general and multi-purpose
dust radiative transfer code. It now includes many advanced techniques
to increase the efficiency, including forced scattering
\citep{1970A&A.....9...53M, 1977ApJS...35....1W}, the peeling-off
technique \citep{1984ApJ...278..186Y}, continuous absorption
\citep{1999A&A...344..282L}, smart detectors
\citep{2008MNRAS.391..617B} and frequency distribution adjustment
\citep{2001ApJ...554..615B, 2005NewA...10..523B}. The code is capable
of producing simulated images, spectral energy distributions,
temperature maps and observed kinematics for arbitrary 3D dusty
systems. SKIRT is completely written in C++ in an object-oriented
programming fashion so it is effortless and straightforward to
implement and use different stellar or dust geometries, dust mixtures,
dust grids, etc.

The default mode in which SKIRT operates is the panchromatic
mode. In this mode, the simulation covers the entire wavelength regime
from UV to millimeter wavelengths and guarantees an energy balance in
the dust, i.e.\ at every location in the system, the emission spectrum
of the dust at infrared and submm wavelengths is determined
self-consistently from the amount of absorbed radiation at
UV/optical/NIR wavelengths. In this mode, the SKIRT code is easily
parallelized to run on shared memory machines using the OpenMP
library: every single thread deals with a different wavelength. This
panchromatic SKIRT code has been used to predict and interpret the
far-infrared emission from a variety of objects, including edge-on
spiral galaxies \citep{2010A&A...518L..39B, 2011ApJS..196...22B,
  2012MNRAS.419..895D}, elliptical galaxies
\citep{2010A&A...518L..54D, 2010A&A...518L..45G}, active galactic
nuclei \citep{2011BaltA..20..490S, 2012MNRAS.420.2756S} and post-AGB
stars \citep{2007BaltA..16..101V}.  

Besides the panchromatic mode, it is also possible to run SKIRT in a
monochromatic mode so it is less time consuming when one wants to
model images at one particular wavelength. In this mode, the OpenMP
parallelization is included by distributing the different photon
packages in the simulation over all available threads, which implies
that some precautions must be taken to avoid race conditions.  This
monochromatic mode is particularly useful when one wants to fit
radiative transfer models to a particular observed image at UV,
optical or near-infrared wavelengths, which we deal with in this
paper.

\subsection{The GAlib genetic algorithm library}
\label{GAlib.sec}

Genetic algorithms are problem solving systems based on evolutionary
principles. In essence evolution theory describes an optimization
process of a population to a given environment. The core difference
between genetic algorithms and most other stochastic methods is that a
genetic algorithm works with a set (population) of possible solutions
(individuals) to the problem (environment). Each individual consists
of a number of parameters (genes). For each of these genes there is a
number of possible values which we call alleles. These alleles do not
have to be a discrete set, but can be defined as a range or pool where
the genes should be drawn from.

The algorithm starts by defining both the size and the content of the
initial population (generation 0). The individuals can be created
randomly from the gene pool or they can be manually defined. Each of the
initial solutions is then evaluated and given a certain ``fitness"
value.  The individuals that meet certain criteria of fitness are then
used to crossover and produce the first offspring (generation
1). Another, more convenient way of determining which individuals are
fit for reproduction is by determining a crossover rate. If, for
example, the crossover rate is set to 0.6, the 60\% best individuals
will be used to produce offspring. We can also define a mutation rate
in a similar way. This rate determines what the odds are for a certain
gene (and not individuals) of the offspring to undergo a random
mutation. In practice this is done by removing one of the gene values
and replacing it with a new possible value. After this step we return
the fitness value. The individuals fit for
reproduction are then selected and a new generation (generation 2) is
created. This cycle is then repeated until a certain fitness value is
obtained or until a pre-defined number of generations is
reached. Since the better individuals of a population are always
preferred, we expect the population average to shift to a better value
where the better individuals are preferred again, etc. 



Genetic algorithms have been applied successfully to a large range of
test problems, and are nowadays widely used as a reliable class of
global optimizers. They are becoming increasingly popular as a tool in
various astrophysical applications. Most astrophysical applications so
far have used the publicly available PIKAIA code
\citep{1995ApJS..101..309C}, originally developed in Fortran 77 and
now available in Fortran 90 and IDL as well.

We have selected the publicly available GAlib library \citep{Wall96}
for our purposes. One of the main reasons for choosing GAlib is that, like the SKIRT radiative transfer code, it is written in
C++ and thus guarantees straightforward interfacing. Moreover, the
library has been properly tested and adapted over the years and has
been applied in many scientific and engineering applications. It comes
with an extensive overview of how to implement a genetic algorithm and
examples illustrating customizations of the GAlib classes. To
illustrate the performance of the GAlib library, we have applied it to
the same test function as first suggested by
\cite{1995ApJS..101..309C} but also investigated by others
\citep{2009A&A...501.1259C, 2012arXiv1202.1643R}. The goal of the
exercise is the find the global maximum of the function
\begin{equation}
  f(x,y)
  =
  \Bigl[
    16\,x\,(1-x)\,y\,(1-y)\sin(n\pi x)\sin(n\pi y)
  \Bigr]^2  
 \label{Charbonneau}
\end{equation}
in the unit square with $x$ and $y$ between 0 and 1. The function is
plotted in Figure~{\ref{Charbonneau.pdf}} for $n = 9$. It is clear
that this function is a severe challenge for most optimization
techniques: the search space contains many steep local maxima
surrounding the global maximum at $(x,y) = (0.5,0.5)$. In addition,
the global maximum is not significantly higher compared to the surrounding local maxima. A classical hill climbing
method would most definitely get stuck in a local maximum. An
iterative hill climbing method, which restarts at randomly chosen 
points, would also be able to solve this problem but it would 
normally take much longer to do so compared to a 
genetic algorithm.

\begin{figure}
\centering
\includegraphics[width=0.48\textwidth]{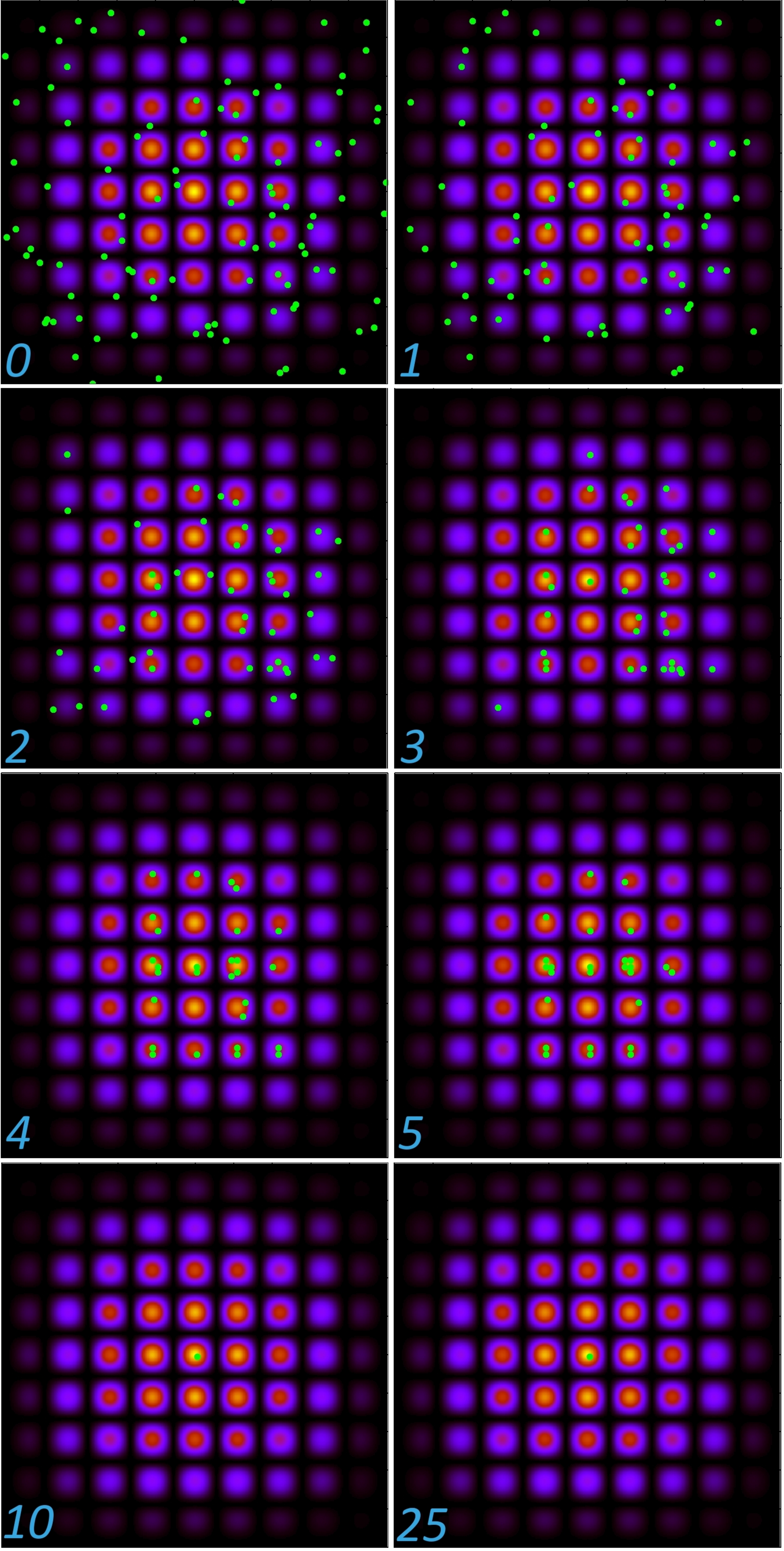}
\caption{Illustration of GAlib's approach to find the global maximum
  of the function~(\ref{Charbonneau}). The different panels show the
  100 individuals of a population 0, 1, 2, 3, 4, 5, 10 and 25,
  characterized by a mutation probability of 0.3\% and a crossover
  rate of 65\%. For the first few generations, the individuals are
  distributed randomly, starting from generation 5 the individuals are
  clearly centered around the local maxima and continue to converge to
  the true maximum.}
  \label{GAlibCharbonneau.pdf}
\end{figure} 

We demonstrate how the GAlib genetic algorithm library behaves in this
complex search space by plotting the contours and overplotting the
position of the individuals for each generation.  The results are
shown in Figure~{\ref{GAlibCharbonneau.pdf}}. For these tests we use
the same values for the mutation probability and crossover rate as
\citet{1995ApJS..101..309C}, namely 0.3\% and 65\% respectively, and
the population size is set to 100. As it is shown, the algorithm is
capable of efficiently determining the maximum. Starting from
generation 5 the individuals are clearly centered around the local
maxima and continue to converge to the true maximum. At generation 10
all individuals are very close to the global maximum, and we note few
changes between generations 10 and 25. This can be explained by the
large population inertia. The large population prohibits the fast
alteration to a favorable mutation. Higher mutation rates could be a
possible way to increase accuracy but we have to keep in mind that
this could come at the cost of losing the global maximum
\citep{1995ApJS..101..309C}. Furthermore it is not our ultimate goal
to optimize this given problem in the best possible manner.

The final result we get using this method after 25 generations is:
$(x,y) = (0.502, 0.498)$ with $f(x,y)=0.994$.  It should be noted that
this solution is not a special case and that the algorithm delivers
this result almost every time. After 1000 consecutive runs we get an
average result
\begin{gather}
  x = 0.501 \pm 0.004 \\
  y = 0.500 \pm 0.004
\end{gather}  
When we increase the number of generations, we obviously still recover
the global maximum and reduce the standard deviation: for a population
of 100 and 100 generations we get
\begin{gather}
  x = 0.500 \pm 0.003 \\
  y = 0.501 \pm 0.003
\end{gather}  
As a final example, we increase the mutation rate to 30\% to show it improves
the accuracy. Again after 100 generations, we now find
\begin{gather}
  x = 0.5000 \pm 0.0002 \\
  y = 0.5000 \pm 0.0002
\end{gather}
Looking at Figure~{\ref{GAlibCharbonneau.pdf}} and comparing with the
result, $f(x,y)=0.978322$ (ranked case, 40 generations), obtained by
\citet{1995ApJS..101..309C}, it can be seen that, for this problem,
GAlib performs at least as good (see the 25 generation case) as the
PIKAIA code and we can be confident to use the GAlib code for our
purposes.\footnote{A possible explanation for this difference might be
  the generational versus the steady state reproduction. It has to be
  noted, however, that the PIKAIA results we quote are from
  \citet{1995ApJS..101..309C}. Many researchers have adapted and
  updated the PIKAIA code since 1995, probably also increasing
  its efficiency.}

\subsection{FitSKIRT}

Our goal is to develop a fitting program that optimizes the parameters
of a 3D dusty galaxy model in such a way that its apparent image on
the sky fits an observed image. This comes down to an optimization
function, where the objective function to be minimized is the $\chi^2$
value,
\begin{equation}
  \chi^2(\boldsymbol{p})
  =
  \sum_{j=1}^{N_{\text{pix}}}
  \left[\frac{I_{\text{mod},j}(\boldsymbol{p}) - I_{\text{obs},j}}
    {\sigma_j(\boldsymbol{p})}\right]^2
\label{chi2}
\end{equation}
In this expression, $N_{\text{pix}}$ is the total number of pixels in
the image, $I_{\text{mod},j}$ and $I_{\text{obs},j}$ represent the
flux in the $j$'th pixel in the simulated and observed image
respectively, $\sigma_j$ is the uncertainty, and $\boldsymbol{p}$
represents the dependency on the parameters of the 3D model
galaxy. Note that, contrary to most $\chi^2$ problems, the uncertainty
$\sigma_j$ in our case depends explicitly on the model: it can be
written as
\begin{equation}
  \sigma_j(\boldsymbol{p})
  =
  \sqrt{\sigma_{\text{obs},j}^2+\sigma_{\text{mod},j}^2(\boldsymbol{p})}
  \label{sigma}
\end{equation}
The factor $\sigma_{\text{obs},j}$ represents the uncertainty on the
flux in the $j$'th pixel of the observed image, usually dominated by a
combination of photon noise and read noise. It can be calculated using
the so-called CCD equation \citep{1981SPIE..290...28M,
  1991PASP..103..122N, 2006hca..book.....H}. The factor
$\sigma_{\text{mod},j}(\boldsymbol{p})$ represents the uncertainty on
the flux in the $j$'th pixel of the simulated image corresponding to
the Monte Carlo simulation with model parameters
${\boldsymbol{p}}$. It can be calculated during the Monte Carlo
radiative transfer simulation according to the recipes from
\citet{2001ApJ...551..269G}.

The full problem we now face is to minimize the $\chi^2$ measure
(\ref{chi2}) by choosing the best value of 
${\boldsymbol{p}}$ in the model parameter space. As the link between
the model image and the model parameters is non-trivial (it involves a
complete radiative transfer simulation), this comes down to a
multidimensional, non-linear, non-differentiable optimization problem.

\begin{figure}
 \centering
 \includegraphics[width=0.45\textwidth]{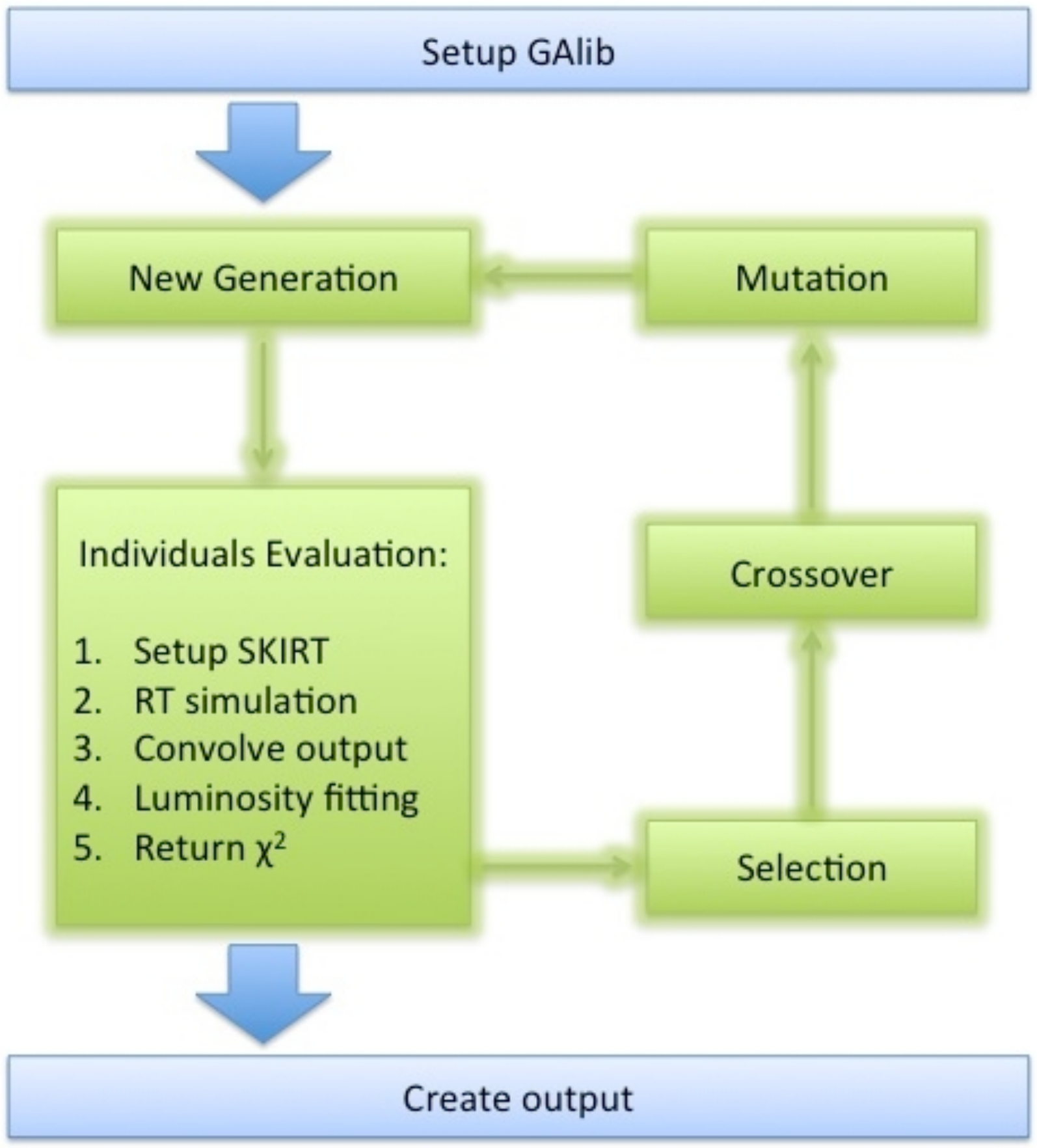} 
 \caption{The main flowchart of the FitSKIRT procedure. Details on the
   different steps are given in the text.}
  \label{FlowChart.pdf}
\end{figure} 

Our approach to this problem resulted in  the FitSKIRT program, a code that
combines the radiative transfer code SKIRT and the genetic algorithm
library GAlib. Figure~{\ref{FlowChart.pdf}} shows a flowchart of the
major parts of the FitSKIRT program.

The first step in the process consists of setting up the ingredients
for the genetic algorithm. This consists in the first place of
defining the reference image and defining the parameterized model
describing the distribution and properties of stars and dust in the
model galaxy. Apart from setting up this model, we also select the
range of the parameters in the model, and define all genetic algorithm
parameters like crossover rate, selection and reproduction scheme,
mutation rate, etc.

In the second phase the genetic algorithm loop (green flow) is
started. The initial population is then created by randomly drawing the
parameter values ${\boldsymbol{p}}$ from the predefined
ranges so the genetic algorithm starts out by being uniformly spread
across the entire parameter space. The next step is to evaluate 
which individuals provide a good fit and which do not. 
This is done by starting a monochromatic SKIRT simulation using the 
parameters ${\boldsymbol{p}}$
defined by the genes of the individual.  Once a simulation is done
we compare the resulting frame and the reference image and give the
corresponding individual an objective score. Between the creation of the
image and returning the actual $\chi^2$ value, the
simulated frame is convolved with the point spread function (PSF) of
the observed image. The third step is then concluded by returning the
final $\chi^2(\boldsymbol{p})$ for this individual. After the entire
population has been evaluated the best models are selected and
offspring is created by crossover. Depending on the mutation rate,
some of the genes of these new individuals will undergo a random
mutation. After this step we obtain our new generation which is
about to be evaluated next. This loop continues until a predefined
number of generations is reached. 

Finally, when the genetic algorithm loop ends, the convolved, best
fitting frame is created again and the residual frame is
determined. These residual frames are useful to investigate which
areas of the references frames are well fitted and which are not. They
can provide additional insight on the validity and consistency of the
models themselves.


In principle, the genetic algorithm searches for the best fitting
model in the entire $N$-dimensional parameter space, where $N$ is the
number of free parameters in the model. In general, the model image,
and hence the $\chi^2$ measure (\ref{chi2}), depends in a complex,
non-linear way on the different parameters in the model, such as
scalelengths, viewing angles, etc. There is one exception however:
the radiative transfer problem is linear with respect to the total
luminosity of the system (which always is one of the parameters in the
model). This allows us to treat the total luminosity separately from
the other parameters, and determine its best value outside the genetic
algorithm minimization routine. This step effectively decreases the
dimensionality of the parameter space the genetic algorithm has to
investigate from $N$ to $N-1$. In practice, this is implemented in the
following way. Assume our 3D model is defined by the $N$ parameters
${\boldsymbol{p}} = (p_1,\ldots,p_{N-1},L_{\text{tot}})$. In every
step of the genetic algorithm, the code selects a set of $N-1$
parameters $(p_1,\ldots,p_{N-1})$ and starts a SKIRT radiative
transfer simulation to create a model image, based on these $N-1$
parameters with a dummy value for $L_{\text{tot}}$, and this model
image is convolved with the observed image PSF. Before the $\chi^2$
value corresponding to this set of parameters is calculated, the code
determines the best fitting total luminosity of the model that
minimizes the $\chi^2$ value (\ref{chi2}) for the particular values of
these $N-1$ parameters. To this aim, we can use a simple
one-dimensional optimization routine. Because no noise is added
between the creation of the image and determining the final $\chi^2$
value of the genome and because the one-dimensional luminosity space
contains no local minima, a fast and simple golden section search is
able to determine the correct value easily.

Apart from finding the parameters that correspond to the model which
best fits the observed image, we also want a measure of the
uncertainties on these parameters. A disadvantage of most stochastic
optimization methods, or more specific genetic algorithms, is that
such a measure is not readily available.  A way of partially solving
this problem is by using a statistical resampling technique like
bootstrapping or jackknifing \citep{1992nrfa.book.....P,
  2010MNRAS.408.1879N, 2012JCAP...11..033N}.  This basically comes
down to iteratively replacing a predefined number of the simulated
points by the actual data points and comparing the resulting objective
function value. Even these estimations of the standards errors should
not be used heedlessly to construct confidence intervals since they
remain subject to the structure of the data.  Bootstrapping also
requires the distribution of the errors to be equal in the data set
and the regression model \citep{Sahinler07, 1992nrfa.book.....P}.
Since this is not entirely the case for this problem, as
equation~(\ref{chi2}) shows, an alternative approach was used to
determine the uncertainty.

Since genetic algorithms are essentially random and the parameter space
is so vast and complex, the fitting procedure can be repeated multiple
times without resulting in the exact same solution. The difficulty of differentiating between some individuals because of
some closely correlated parameters will be reflected in the final
solutions.  The entire fitting procedure used here consists
of running five independent FitSKIRT simulations, all with the same
optimization parameters.  The standard deviation on these five
solutions is set as an uncertainty on the "best" solution (meaning the
lowest objective function value). While computationally expensive,
this method still allows for better solutions to be found.  The spread gives some insight in how well the fitting procedure
is able to constrain some parameters and which parameters correlate.
When the solutions are not at all coherent, this can also indicate
something went wrong during the optimization process (f.e. not enough
individuals or evaluated generations).

\section{Application on test images}
\label{TestImages.sec}

In this section, we apply the FitSKIRT code on a simulated test image,
in order to check the accuracy and effectiveness of this method in
deriving the actual input parameters. 

\subsection{The model}
\label{Model.sec}

The test model consists of a simple but realistic model for a dusty
edge-on spiral galaxy, similar to the models that have been used to
actually model observed edge-on spiral galaxies
\citep[e.g.][]{1987ApJ...317..637K, 1997A&A...325..135X,
  1999A&A...344..868X, 2007A&A...471..765B, 2008A&A...490..461B,
  2010A&A...518L..39B, 2011A&A...527A.109P, 2011ApJ...741....6M,
  2012A&A...541L...5H}. The system consists of a stellar disc, a
stellar bulge and a dust disc.

The stellar disc is characterized by a double-exponential disc,
described by the luminosity density
\begin{equation}
  j(R,z)
  =
  \frac{L_{\text{d}}}{4\pi\,h_{R,*}^2\,h_{z,*}}
  \exp\left(-\frac{R}{h_{R,*}}\right)
  \exp\left(-\frac{|z|}{h_{z,*}}\right)
\end{equation}
with $L_{\text{d}}$ the disc luminosity, $h_{R,*}$ the radial scalelength and
$h_{z,*}$ the vertical scaleheight. For the stellar bulge, we assume
the following 3D distribution,
\begin{equation}
  j(R,z)
  =
  \frac{L_{\text{b}}}{q\,R_{\text{e}}^3}\,
  {\cal{S}}_n\left(\frac{m}{R_{\text{e}}}\right)
\end{equation}
where 
\begin{equation}
  m = \sqrt{R^2+\frac{z^2}{q^2}}
\end{equation}
is the spheroidal radius and ${\cal{S}}_n(s)$ is the S\'ersic
function, defined as the normalized 3D luminosity density
corresponding to a S\'ersic surface brightness profile, i.e.\ 
\begin{gather}
  {\cal{S}}_n(s)
  =
  -\frac{1}{\pi}
  \int_s^\infty
  \frac{{\text{d}}I}{{\text{d}}t}(t)\,
  \frac{{\text{d}}t}{\sqrt{t^2-s^2}}
  \\
  I(t)
  =
  \frac{b^{2n}}{\pi\,\Gamma(2n+1)}
  \exp\left(-b\,t^{1/n}\right)
\end{gather}
This function can only be expressed in analytical form using special
functions \citep{2002A&A...383..384M, 2011A&A...525A.136B,
  2011A&A...534A..69B}. Our bulge model has four free parameters: the
bulge luminosity $L_{\text{b}}$, the effective radius $R_{\text{e}}$,
the S\'ersic index $n$ and the intrinsic flattening $q$.

The dust in the model is also distributed as a double-exponential
disc, similar to the stellar disc,
\begin{equation}
  \rho_{\text{d}}(R,z)
  =
  \frac{M_{\text{d}}}{4\pi\,h_{R,{\text{d}}}^2\,h_{z,{\text{d}}}}
  \exp\left(-\frac{R}{h_{R,{\text{d}}}}\right)
  \exp\left(-\frac{|z|}{h_{z,{\text{d}}}}\right)
\end{equation}
with $M_{\text{d}}$ the total dust mass, and $h_{R,{\text{d}}}$ and
$h_{z,{\text{d}}}$ the radial scalelength and vertical scaleheight of
the dust respectively. The central face-on optical depth, often used
as an alternative to express the dust content, is easily calculated
\begin{equation}
  \tau_{\text{f}}
  \equiv
  \int_{-\infty}^{\infty}
  \kappa\,\rho(0,z)\,{\text{d}}z
  =
  \frac{\kappa\,M_{\text{d}}}{2\pi\,h_{R,{\text{d}}}^2}
\label{tauf}
\end{equation}
where $\kappa$ is the extinction coefficient of the dust.

\begin{table*}
  \centering
  \begin{tabular}{|c|c|c|c|c|c|}
    \hline
    & & & & & \\
    parameter & unit & reference & 1 parameter & 3 parameters & 11
    parameters \\ 
    & & & & & \\ \hline 
    & & & & & \\
    $M_{\text{d}}$ & $10^7~M_\odot$ & 4 & $4.00\pm0.02$ &
    $3.98\pm0.08$ & $3.75\pm0.5$ \\
    $\tau_{\text{f}}$ & -- & 0.80 & 0.80 $\pm$ 0.01 & 0.86 $\pm$ 0.08 & $0.98 \pm 0.14$ \\ 
    $h_{R,{\text{d}}}$ & kpc & 6.6 & -- & $6.50\pm0.40$ & $5.83\pm0.75$ \\
    $h_{z,{\text{d}}}$ & pc & 250 & -- & $241\pm16$ & $245\pm30$ \\
    $L_{\text{tot}}$ &$10^{9}\ L_{\odot}$ & 1 & -- & -- & $0.97 \pm 0.09$ \\
    $h_{R,*}$ & kpc & 4.4 & -- & -- & $4.36\pm0.26$ \\
    $h_{z,*}$ & pc & 500 & -- & -- & $519\pm51$ \\
    $B/D$ & -- & 0.33 & -- & -- & $0.34\pm0.08$ \\ 
    $R_{\text{e}}$ & kpc & 2 & -- & -- & $1.75\pm0.24$ \\
    $n$ & -- & 2.5 & -- & -- & $2.5\pm0.5$ \\ 
    $q$ & -- & 0.5 & -- & -- & $0.50\pm0.04$ \\
    $i$ & deg & 89 & -- & -- & $88.9\pm0.1$ \\ 
    & & & & & \\
    \hline 
  \end{tabular}
  \vspace*{2ex}
  \caption{Input values and fitted values of the parameters of the
    model used in the test simulations in Section~{\ref{TestImages.sec}}. The values used
    to create the reference image (see Figure~{\ref{TestImage.pdf}})
    are given in the third column. The fourth, fifth and sixth column
    list the fitted values for these
    parameters, together with their 1$\sigma$ error bars, for the fits
    with one, three and eleven free parameters respectively.}
  \label{parameters.tab}
\end{table*}

In total, this 3D model has 10 free parameters, to which the
inclination $i$ of the galaxy with respect to the line of sight should
be added as an 11th parameter. To construct our reference model, we
selected realistic values for all parameters, based on average
properties of the stellar discs, bulges and dust discs in spiral
galaxies \citep{1999A&A...344..868X, 2002MNRAS.334..646K,
  2004A&A...414..905H, 2007A&A...471..765B, 2012A&A...540A..52C}. The
set of parameters is listed in the third column of
Table~{\ref{parameters.tab}}. We constructed a V-band image on which
we will test the FitSKIRT code by running the SKIRT code in
monochromatic mode on this model, fully taking into account absorption
and multiple anisotropic scattering. The optical properties of the
dust were taken from \citet{2007ApJ...657..810D}. The resulting input
image is shown in the top panel of Figure~{\ref{TestImage.pdf}}. It is
$500 \times 100$ pixels in size, has a pixel scale of 100 pc/pixel and
the reference frame was created using $10^7$ photon packages.

We now apply the FitSKIRT code with the same parameter model to
this artificial test image and investigate whether we can recover the
initial parameters of the model. We proceed in different steps, first
keeping a number of parameters in the model fixed to their input value
and gradually increasing the number of free parameters. 

\subsection{One free parameter}

\begin{figure}
\centering
\includegraphics[width=0.45\textwidth]{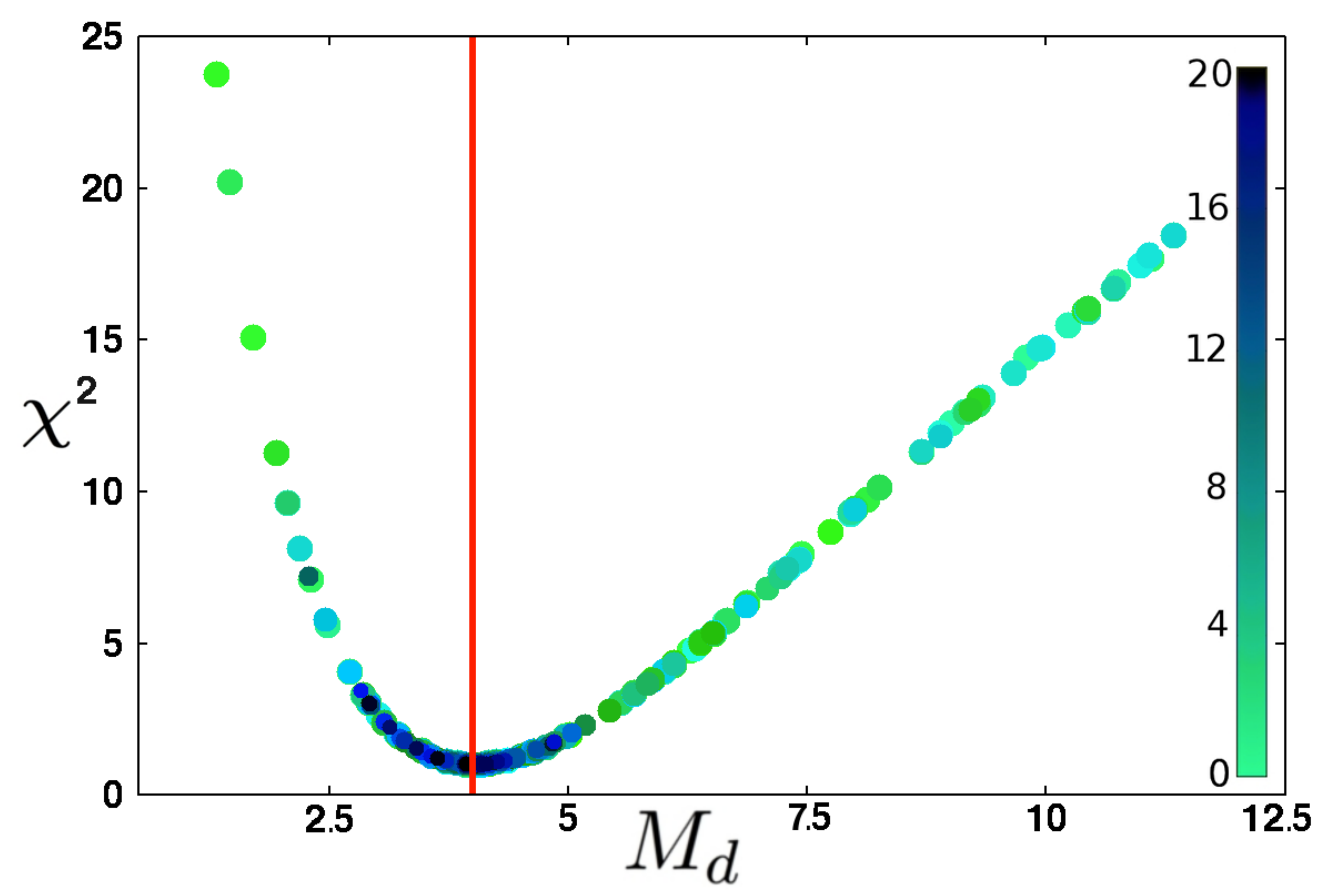} 
\caption{Evolution of the determination of the dust mass in the
  FitSKIRT test simulation with one free parameter. The plot shows the
  distribution of the $\chi^2$ value for each of the 100 individuals
  for 20 generations. The colors represent the different generations
  and the input value is indicated by a red line.}
 \label{OneParameter.pdf}
\end{figure}

\begin{figure*}
  \centering 
  \includegraphics[width=0.8\textwidth]{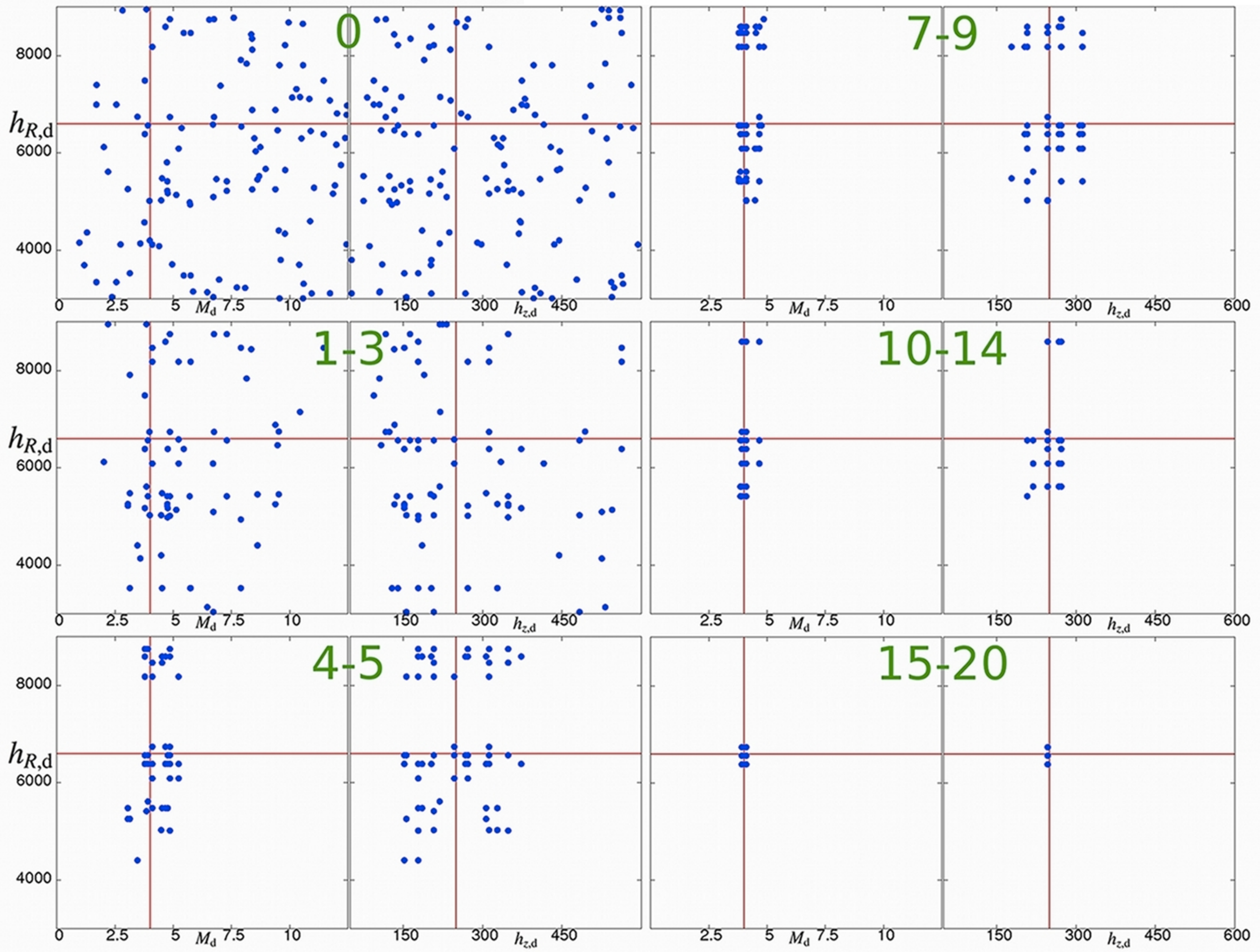} 
  \caption{Evolution of the dust parameters for a population of 100
    individuals for the test model with three free parameters. The
    different panels show the position of each individual in the
    $h_{R,{\text{d}}}$ versus $h_{z,{\text{d}}}$ and the
    $h_{R,{\text{d}}}$ versus $M_{\text{d}}$ projections of the 3D
    parameter space, for different generations (indicated by the green
    numbers on the top of each panel). The input values of the
    parameters are indicated by red lines.}
  \label{ThreeParameters.pdf}
\end{figure*}

To do some first basic testing with FitSKIRT, no luminosity fitting is
used on any of the images.  Apart from the dust mass $M_{\text{d}}$
all parameters are fixed and set to the same of the simulated
image. We let FitSKIRT search for the best fitting model with the dust
mass ranging between $5\times10^6$ and $1.25\times10^8~M_\odot$. Note
that this interval is not symmetric with respect to the model input
value of $4\times10^7~M_\odot$, which makes it slightly more difficult
to determine the real value.

Figure~{\ref{OneParameter.pdf}} shows the evolution of a population of
100 individuals through 20 generations with a mutation probability of
10\% and a crossover rate of 70\%. In fact, when using genomes with
only one free parameter, the crossover rate becomes quite
meaningless. This is because the crossover between two different
genomes will result in an offspring that is essentially exactly the
same as the best of the parent genomes. This duplication will,
however, still result in a faster convergence since the mutation
around those genomes will generally be in a better area than around a
random position. FitSKIRT recovers a dust mass of
$(4.00\pm0.02)\times10^7~M_\odot$, exactly equal to the dust mass of
the input model. This figure shows that the parameter space is indeed
asymmetrical around the true value (indicated by the red line) and
that the entire population is gradually shifting towards the optimal
value.

\subsection{Three free parameters}
\label{ThreeFree.sec}

\begin{figure*}
  \centering
  \includegraphics[width=0.7\textwidth]{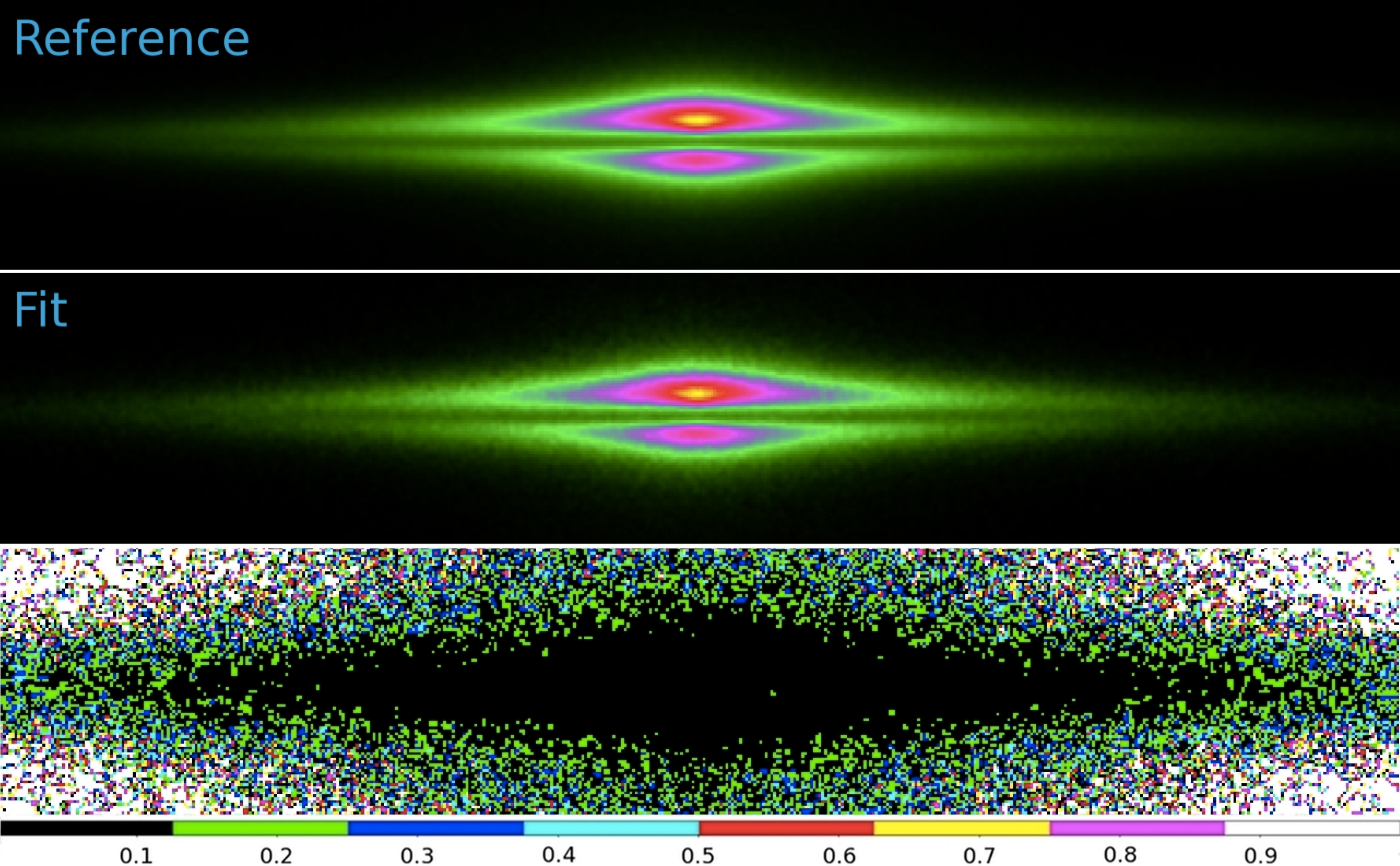}
  \caption{Results of the FitSKIRT radiative transfer fit to the
    artificial edge-on spiral galaxy with 11 free parameters. The
    reference input V-band image is shown on the top panel, the
    FitSKIRT solution is shown in the middle panel, and the bottom
    panel represents the residual frame.}
  \label{TestImage.pdf}
\end{figure*} 

As a second step we considere the case where we fitted three free
parameters, more specifically the three parameters defining the dust
distribution: the dust mass $M_{\text{d}}$, the radial scale length
$h_{R,{\text{d}}}$ and vertical scale height $h_{z,{\text{d}}}$. The
dust parameters are hard to determine individually since they have to
be determined directly from the dust lane. The stellar parameters on
the other hand are more easily determined and constrained in a larger
region outside the dust lane. The problem is close to being degenerate
when we look at the dust scale length and dust mass in exact edge-on,
since changing them will roughly affect the same pixels.

We consider a wide possible range for the free parameters: the dust
scale length was searched between 3 and 9 kpc, the scale height
between 50 and 600 pc and for the dust mass we considered a range
between $5\times10^6$ and $1.25\times10^8~M_\odot$.  In
Figure~{\ref{ThreeParameters.pdf}} we can see the evolution of a
population consisting of 100 individuals through 20 generations in the
$h_{R,{\text{d}}}$ versus $h_{z,{\text{d}}}$ and the
$h_{R,{\text{d}}}$ versus $M_{\text{d}}$ projections of the 3D
parameter space.  For now we only consider a crossover of 70$\%$ and
disable the mutation so we can take a closer look at the global
optimization process. As we can see the entire population slowly
converges until, in the end, it is located entirely around the actual
value. This is indeed what we would expect from a global optimization
process. The final best fit values are summarized in the fifth column
of Table~{\ref{parameters.tab}}. With only three free parameters, FitSKIRT is still able to converge towards the exact
values within the uncertainties. Notice that out of all parameters, the
dust scale length $h_{R,d}$ is the hardest to constrain for edge-on
galaxies.  The values can be constrained even more if we use the
mutation operator because the search space is then investigated in a
multi-dimensional normal distribution around these individuals. Since
in the end most individuals reside in the same area, this area will be
sampled quite well.

\subsection{Eleven free parameters}
\label{ElevenFree.sec}

As a final test, FitSKIRT was used to fit all 11 parameters of the
model described in section~{\ref{Model.sec}}. This corresponds to a
full 10+1 dimensional optimization problem (10 parameters fitted by
the genetic algorithm and the total luminosity determined
independently by a golden section search). { The appendix section also contains a comparison between genetic algorithms, Levenberg-Marquardt and downhill-simplex method for this specific case.} In order to accommodate the
significant increase of the parameter space, we also changed the
parameters of the genetic algorithm fit: we consider 100 generations
of 250 individuals each and set the mutation rate to 5\% and the
crossover rate to 60\%. We also have to keep in mind that larger
populations are less sensitive to noise \citep{Goldberg91}.

The result of this fitting exercise is given in the sixth column of
Table~{\ref{parameters.tab}}. These results are also shown graphically
in Figure~{\ref{TestImage.pdf}}. The central panel shows the simulated
image of our best fitting model, to be compared with the reference
image shown on the top panel. The bottom panel gives the residual
image. This figure clearly shows that the reference image is
reproduced quite accurately by FitSKIRT and the residual frame shows a
maximum of $10\%$ discrepancy in most of the pixels, which is quite
impressive considering the complexity of the problem and vastness of
the parameter space. Apart from a good global fit, the individual parameters are also very well recovered: we can recover all fitted parameters within the $1\sigma$ uncertainty interval (only the bulge
effective radius is just outside this one standard deviation
range). The uncertainty estimates are also useful to get more insight
in which parameters are easily constrained and which are not. From
Table~{\ref{parameters.tab}} it is clear that the dust scale length
and dust mass are the hardest to constrain. These parameters are close to degenerate when fitting edge-on spiral galaxies as the fitting routine is most sensitive to the edge-on optical depth, i.e.\ the dust column within the plane of the galaxy. Both a dust distribution with a large dust mass but a small scale length and a
distribution with a small dust mass and a large scale length can
conspire to give similar edge-on optical depths.

\section{Application on observed data: NGC\,4013}
\label{NGC4013.sec}

In this final section we apply FitSKIRT to recover the intrinsic
distribution of stars and dust in the edge-on spiral galaxy
NGC\,4013. Located at a distance of about 18.6~Mpc
\citep{1997ApJS..109..333W, 2002ApJ...565..681R, 2009AJ....138..323T}
and with a $D_{25}$ diameter of 5.2~arcmin, this galaxy is one of the
most prominent edge-on spiral galaxies. It is very close to exactly
edge-on and its dust lane can be traced to the edges of the
galaxy. Conspicuous properties of this galaxy are its box- or
peanut-shaped bulge \citep{1982ApJ...256..460K, 1986AJ.....91...65J,
  1999A&A...343..740G} and the warp in the gas and stellar
distribution \citep{1987Natur.328..401B, 1991A&A...242..301F,
  1995A&A...295..605B, 1996A&A...306..345B, 2009MNRAS.396..409S}. The
main reason why we selected this particular galaxy is that it has been
modelled before twice using independent radiative transfer fitting
methods, namely by \citet{1999A&A...344..868X} and
\citet{2007A&A...471..765B}. This allows a direct comparison of the
FitSKIRT algorithm with other fitting procedures in a realistic
context.

\begin{figure*}
  \centering
  \includegraphics[width=\textwidth]{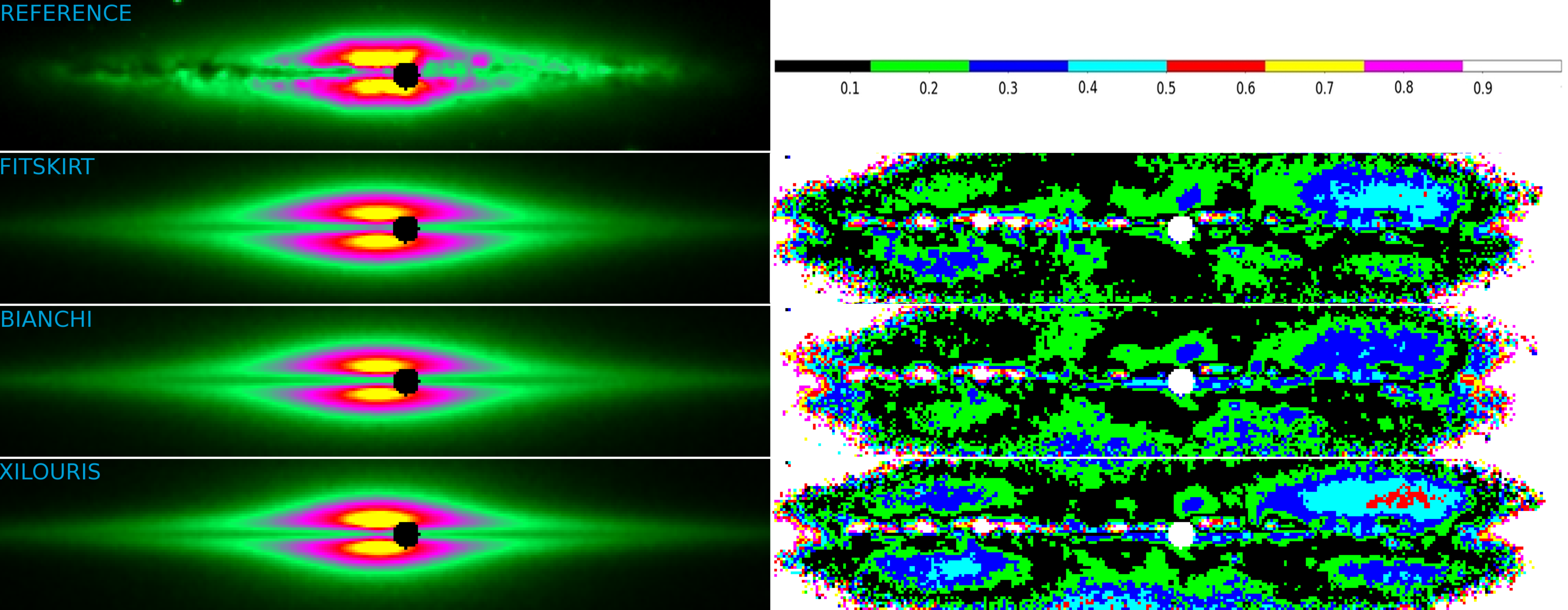}
  \caption{Radiative transfer model fits to a V-band image of
    NGC\,4013. The top right panel shows the observed image, the
    subsequent rows show the best fitting SKIRT image, and the models
    by \citet{2007A&A...471..765B} and
    \citet{1999A&A...344..868X}. The panels on the right-hand side
    show the residual images corresponding to these models.}
  \label{NGC4013.pdf}
\end{figure*}

We use a V-band image of NGC\,4013, taken with the Telescopio
Nazionale Galileo (TNG). Full details on the observations and data
reduction can be found in \citet{2007A&A...471..765B}. The image can
be found in the top left panel of Figure~{\ref{NGC4013.pdf}}. We apply
the FitSKIRT program to reproduce this V-band image, in a field of
approximatively $5.5' \times 1'$. The same model is used as the test
simulations discussed in Section~{\ref{TestImages.sec}}, i.e.\ a
double-exponential disc and a flattened S\'ersic bulge for the stellar
distribution and a double-exponential disc for the dust
distribution. We do not fix any parameters, i.e.\ we are again facing
a 10+1 dimensional optimization problem. The genetic algorithm
parameters are the same as for the test problem, i.e.\ 100 generations
of 250 individuals each, a mutation rate of 5\% and a crossover rate
of 60\%.

\begin{table}[t!]
  \centering
  \begin{tabular}{|c|c|c|c|c|}
    \hline 
    & & & & \\
    parameter & unit & FitSKIRT & B07 & X99 \\ 
    & & & & \\ \hline 
    & & & & \\
    $\tau_{\text{f}}$ & -- & $0.97\pm0.36$ & 1.46 & $0.67\pm0.01$ \\ 
    $\tau_{\text{e}}$ & -- & $15.1\pm3.1$ & 21.0 & 12.6 \\
    $M_{\text{d}}$ & $10^6~M_\odot$ & $9.9\pm1.9$ & 7.3 & 7.3 \\ 
    $h_{R,{\text{d}}}$ & kpc & $3.00\pm0.46$ & 2.08 & $3.09\pm0.13$ \\
    $h_{z,{\text{d}}}$ & pc & $192\pm16$ & 145 & $164\pm13$ \\
    $h_{R,*}$ & kpc & $2.12\pm0.11$ & 2.89 & $2.45\pm0.13$ \\
    $h_{z,*}$ & pc & $287\pm104$ & 376 & $278\pm13$ \\
    $B/D$ & -- & $1.18\pm0.32$ & 2.13 & 1.47 \\
    $R_{\text{e}}$ & kpc & $2.30\pm0.46$ & 1.91 & $1.79\pm0.05$ \\
    $n$ & -- & $3.2\pm0.7$ & 4 & 4 \\
    $q$ & -- & $0.35\pm0.03$ & 0.37 & $0.44\pm0.01$ \\
    $i$ & deg & $90.0\pm0.1$ & 89.9 & $89.7\pm0.1$ \\
    & & & & \\
    \hline 
  \end{tabular}
  \vspace*{2ex}
  \caption{Parameters of the intrinsic distribution of stars and dust
    in NGC\,4013 as obtained by FitSKIRT, by
    \citet{2007A&A...471..765B} and by
    \citet{1999A&A...344..868X}. For the latter two models, the values
    are scaled to our assumed distance of
    18.6~Mpc and dust masses are calculated using the same values for
    the dust extinction coefficient $\kappa$ as used in FitSKIRT.}
  \label{NGC4013.tab}
\end{table}

\begin{figure}
 \includegraphics[width=0.45\textwidth]{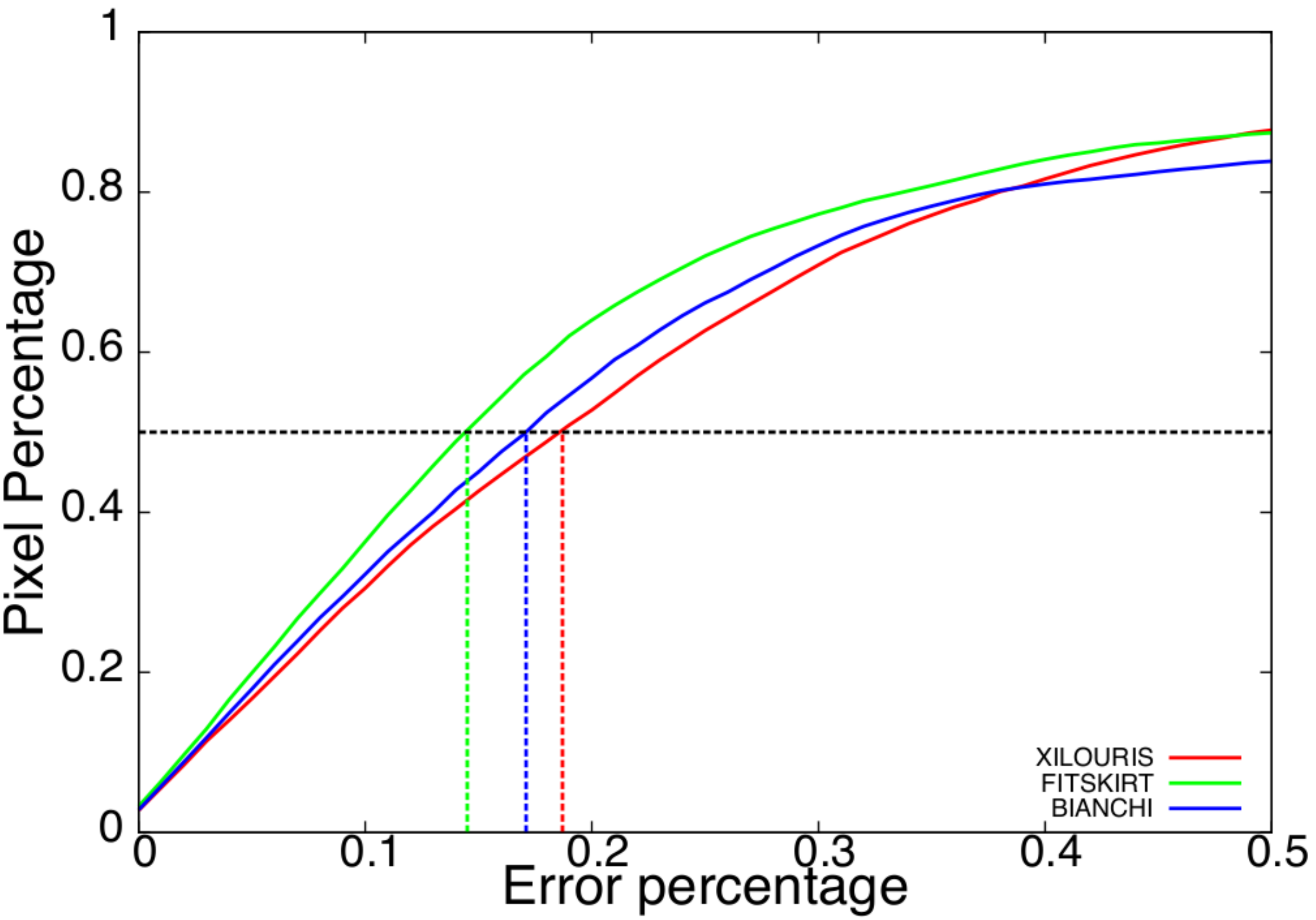} 
 \caption{The cumulative distribution of the number of pixels with a
   residual smaller than a given percentage for the three model fits
   to the V-band image of NGC\,4013. The green, red and blue solid
   lines correspond to the FitSKIRT model, the
   \citet{1999A&A...344..868X} model and the
   \citet{2007A&A...471..765B} model respectively. The dashed
   horizontal line corresponds to half of the total number of pixels.}
 \label{NGC4013-Residuals.pdf}
\end{figure} 

The results of our FitSKIRT fit are given in the third column of
Table~{\ref{NGC4013.tab}}. We find a V-band face-on optical depth of
almost unity, corresponding to a dust mass of
$9.9\times10^6~M_\odot$. In Figure~{\ref{NGC4013.pdf}} we plot the
best fit to the observed V-band image of NGC\,4013 and its
residuals. These two panels show that our FitSKIRT model provides a
very satisfactory fit to the observed image: the residuals between the
image and fit are virtually everywhere smaller than 20\%.  The main
exceptions where the fit is less accurate are the central-left region
of the disc which contains discernible clumpy structures (which are
obviously not properly described by our simple analytical model) and
the top-right region, which is due to the warping of the disc in
NGC\,4013. The quality of our FitSKIRT fit is quantified in
Figure~{\ref{NGC4013-Residuals.pdf}}, where we plot the cumulative
distribution of the residual values of the innermost 15,000 pixels in
the fit, i.e.\ the number of pixels with a residual smaller than a
certain percentage. We see that half of the pixels have a residual
less than 15\%, and almost 90\% of the pixels have a residual less
than 50\%.

In the last two columns of Table~{\ref{NGC4013.tab}} we list the
parameters found by \citet{1999A&A...344..868X} and
\citet{2007A&A...471..765B}, scaled to our assumed distance of
18.6~Mpc and considering the same value for the dust extinction
coefficient $\kappa$, necessary to convert optical depths to dust
masses as in equation~(\ref{tauf}). In Figure~{\ref{NGC4013.pdf}} we
also compare the images and their residuals for the models by
\citet{1999A&A...344..868X} and \citet{2007A&A...471..765B} with the
FitSKIRT model. We reconstructed these models by taking the parameters
from Table~{\ref{NGC4013.tab}} and build a V-band model image of the
galaxy with SKIRT. The same quantitative analysis on the residual
frames as for the FitSKIRT solution is also plotted in
Figure~{\ref{NGC4013-Residuals.pdf}}.

Looking at the model fits and their residuals, it is immediately clear
that the three models are very similar. The largest deviations from
the observed image are found at the same locations for the three
models and hence seem to be due to impossibility of our smooth
analytical model to represent the true distribution of stars and dust
in the galaxies rather than due to the fitting
techniques. Figure~{\ref{NGC4013-Residuals.pdf}} seems to suggest that
the FitSKIRT solution provides a slightly better overall fit to the
image, but this judgement might be biased by the fact that we
reproduced the other models with our SKIRT code (their model might be
slightly different). Looking at the model parameters in
Table~{\ref{NGC4013.tab}}, it is clear that the most prominent
differences between the three models are the different values of the
face-on optical depth and the dust scale length. This finding is not
surprising given that we already found in
Section~{\ref{ThreeFree.sec}} and {\ref{ElevenFree.sec}} that these
parameters are the hardest to constrain. This also corresponds to what
\citep{2007A&A...471..765B} states as the reason for the high optical
depth: the large optical depth is compensated by the smaller scale 
length of the dust disk.

An important aspect to take into account when comparing the results
from the three models is the slight variations in the model setup and
the strong differences in the radiative transfer calculations and the
fitting techniques. \citet{1999A&A...344..868X} used an approximate
analytical method to solve the radiative transfer, based on the
so-called method of scattered intensities pioneered by
\citet{1937ApJ....85..107H} and subsequently perfectionized and
applied by \citet{1987ApJ...317..637K}, \citet{1997A&A...325..135X,
  1998A&A...331..894X, 1999A&A...344..868X},
\citet{2001MNRAS.326..722B}, \citet{2000A&A...362..138P,
  2011A&A...527A.109P} and \citet{2006A&A...456..941M}. To do the
fitting, they used a Levenberg-Marquardt algorithm. They used an
approximation of a de Vaucouleurs model to fit the bulge and excluded
the central 200~pc from the fit. The modelling of NGC\,4013 by
\citet{2007A&A...471..765B} on the other hand was based on the Monte
Carlo code TRADING \citep{1996ApJ...465..127B, 2000A&A...359...65B,
  2008A&A...490..461B}. For the actual fitting, he used a combination
of the Levenberg-Marquardt algorithm (applied to model without
scattering) and the amoeba downhill simplex algorithm. He also used an
approximation of a de Vaucouleurs model to fit the bulge. Given these
differences from the FitSKIRT approach, the agreement between the
parameters of the three models is very satisfactory.

\section{Discussion and conclusions}
\label{Discussion.sec}

In this paper, we have presented the FitSKIRT code, a tool designed to
recover the spatial distribution of stars and dust from fitting
radiative transfer models to UV, optical or near-infrared images. It
combines a state-of-the-art Monte Carlo radiative transfer code with
the power of genetic algorithms to perform the actual fitting. The
noise handling properties together with the ability to uniformly
investigate and optimize a complex parameter space, make genetic
algorithms an ideal candidate to use in combination with a Monte Carlo
radiative transfer code.  Using a standard but challenging
optimization problem, we have demonstrated that the specific genetic
algorithm library chosen for FitSKIRT, GAlib, is reliable and
efficient. We could overcome the main shortcoming of the genetic
algorithm approach, the lack of appropriate error bars on the model
parameters, by deriving the spread of the individual parameters when
applying the optimization process several times with different initial
conditions.

The FitSKIRT program was tested on an artificial reference image of a
dusty edge-on spiral galaxy model created by the SKIRT radiative
transfer code. This reference image was fed to the FitSKIRT code in a
series of tests with an increasing number of free parameters. The
reliability of the code was evaluated by investigating the residual
frames as well as the recovery of the input model parameters. From
both the one and three parameters fittings we concluded that the
optimization process is stable enough and does not converge too fast
towards local extrema . FitSKIRT recovered the input parameters very
well, even for the full problem in which all 11 parameters of the
input model were left unconstrained.

As a final test, the FitSKIRT method was applied to determine the
physical parameters of the stellar and dust distribution in the
edge-on spiral galaxy NGC\,4013 from a single V-band image. Looking at
the cumulative distribution of the number of pixels in the residual
map, we found that FitSKIRT was able to fit half of the pixels with a
residual of less than 15\% and almost 90\% of the pixels with a
residual of less than 50\%. Given that the image of NGC\,4013 clearly
shows a number of regions that cannot be reproduced by a smooth model
(due to obvious clumping in the dust lane and a warp in the stellar
and gas distribution), these statistics are very encouraging. The
values of the fitted parameters and the quality of the fit were compared
with similar but completely independent radiative transfer fits done
by \cite{1999A&A...344..868X} and \cite{2007A&A...471..765B}. There
are some deviations between the results, and we argue that these can
be explained by the degeneracy between the dust scale length and
face-on optical depth, and the differences in the model setup,
radiative transfer technique and optimization strategy. In general,
the agreement between the parameters of the three models is very
satisfactory.

We have demonstrated in this paper that the FitSKIRT code is capable
of recovering the intrinsic parameters of the stellar and dust
distribution of edge-on spiral galaxies by fitting radiative transfer
models to an observed optical image. A future extension we foresee is
to include images at different wavelengths in our modelling
procedure. In its most obvious form, one could run FitSKIRT
independently on different images, and use the results obtained at
different wavelengths to study the wavelength dependence of the stars
and dust \citep{1997A&A...325..135X, 1998A&A...331..894X,
  1999A&A...344..868X, 2007A&A...471..765B}. In particular, the
wavelength dependence of the optical depth, i.e.\ the intrinsic
extinction curve, is a strong diagnostic for the size distribution and
the composition of the dust grains \citep{1983ApJ...272..563G,
  1990A&A...237..215D, 1997A&A...323..566L, 2001ApJ...548..296W,
  2003ApJ...588..871C, 2004ApJS..152..211Z}. One step beyond this is
an expansion of FitSKIRT in which images at different wavelengths are
fitted simultaneously, and a single model is sought that provides the
best overall fit to a set of UV/optical/NIR images. A so-called
{\em{oligochromatic}} radiative transfer fitting could disentangle
some of the degeneracies which monochromatic fitting procedures have
to deal with, and can also be used to predict the corresponding
FIR/submm emission of the system \citep{2000A&A...362..138P,
  2010A&A...518L..39B, 2012arXiv1209.2636D, 2012MNRAS.419..895D}.

The main goal of the paper was to present the philosophy and
ingredients behind the FitSKIRT code, and to demonstrate that it is
capable of determining the structural parameters of the stellar and
dust distribution in edge-on spiral galaxies using UV/optical/NIR
imaging data. Obviously, we have the intention to also apply this code
to real data to investigate the physical properties of galaxies. As
the optimization process in FitSKIRT can cover a large parameter space
and the code is almost fully automated (which implies minimal human
intervention and hence bias), it is a very versatile tool and is ready
to be applied to a variety of galaxies, including larger sets of
observational data.

We are currently applying the code to a set of large edge-on spiral
galaxies in order to recover the detailed distribution of stars and
dust. Our main motivation is to investigate whether the amount of dust
observed in the UV/optical/NIR window agrees with the dust masses
derived from MIR/FIR/submm observations, which has been subject of
some debate in recent years \citep{2001A&A...372..775M,
  2006A&A...459..113M, 2004A&A...425..109A, 2008A&A...490..461B,
  2010A&A...518L..39B, 2011A&A...527A.109P,
  2011ApJ...741....6M}. Major problems that have hampered much
progress in this topic in the past were the limited wavelength
coverage and sensitivity of the FIR/submm observations and the small
number of galaxies for which such observations and detailed radiative
transfer models were available. These problems are now being cured. On
the one hand, we now have a powerful tool available to fit optical/NIR
images in an automated way without the need for and the bias from
human intervention. On the other hand, Spitzer and Herschel have now
provided us with deep imaging observations of significant numbers of
nearby edge-on spiral galaxies at various FIR/submm wavelengths
\citep[e.g.][]{2011A&A...531L..11B, 2012A&A...541L...5H,
  2012arXiv1209.2636D, 2012MNRAS.419..895D, 2012arXiv1204.4726C,
  Verstappen2012}.

We have focused our attention here on edge-on spiral galaxies, which
are the obvious and most popular candidates for radiative transfer
modelling because of their prominent dust lanes. However, these are
not the only targets on which FitSKIRT could be applied. In principle,
the fitting routine can be applied on any geometry, but for
monochromatic fitting the dust parameters can only be constrained for
galaxies with a clear and regular signature of dust extinction. Apart
from edge-on spiral galaxies, an interesting class of galaxies are
early-type galaxies which often show regular and large-scale dust
lanes \citep[e.g.][]{1978ApJ...226L.115B, 1981MNRAS.196..747H,
  1985AJ.....90..183E, 2007A&A...461..103P, 2008MNRAS.390..969F,
  2010MNRAS.409..727F}. We plan to apply our FitSKIRT modelling in the
future to a sample of dust-lane early-type galaxies, primarily
focusing on those galaxies that have been mapped at FIR/submm
wavelengths \citep[e.g.][]{2012ApJ...748..123S}.

Other applications are possible and the authors welcome
suggestions from interested readers.

\section*{Acknowledgements}

We thank Karl Gordon, J\"urgen Steinacker, Ilse De Looze, Joris
Verstappen, Simone Bianchi and Manolis Xilouris, for stimulating
discussions on radiative transfer modelling. We would like to express
our gratitude to Matthew Wall for the creation and sharing of the
GAlib library. GDG is supported by the Flemish Fund for Scientific
Research (FWO-Vlaanderen). JF, MB and PC acknowledge the financial
support of the Belgian Science Policy Office.

\bibliographystyle{aa}
\bibliography{FitSKIRT}
{%
\begin{appendix}
\section{Other optimization techniques}

\begin{figure*}
  \centering
 \includegraphics[width=0.7\textwidth]{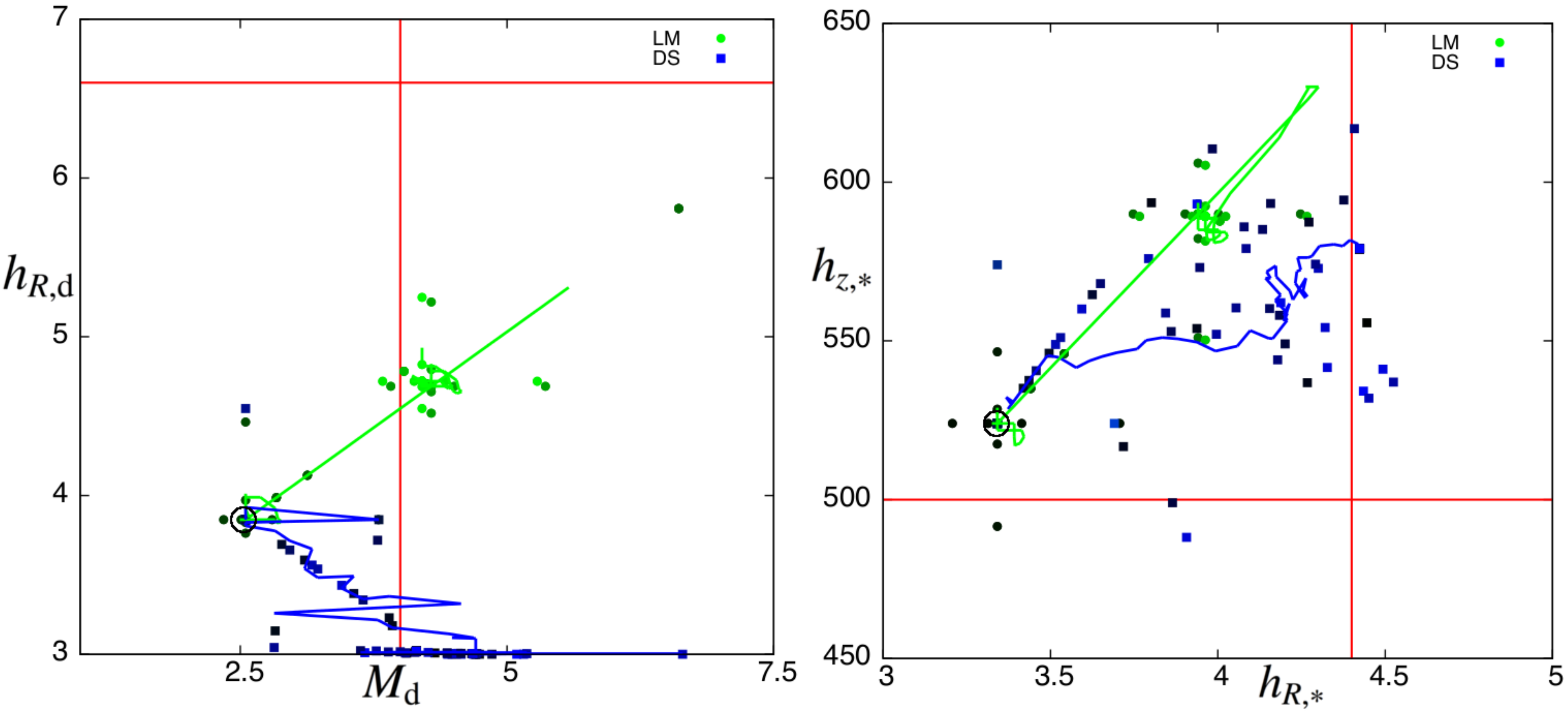} 
 \caption{The function evaluations for both the Levenberg-Marquardt
   and downhill simplex method ranging from black (initial) to blue
   (DS) or green (LM) and the average path. The red lines indicate the
   values used to create the reference frame
   (Figure~{\ref{TestImage.pdf}}). Left: the variation of the dust
   mass and the scale length of the dust disk. Right: the variation of
   the scale length and scale height of the stellar disk.}
 \label{D_S2.pdf}
\end{figure*} 

\begin{figure*}
  \centering
  \includegraphics[width=0.7\textwidth]{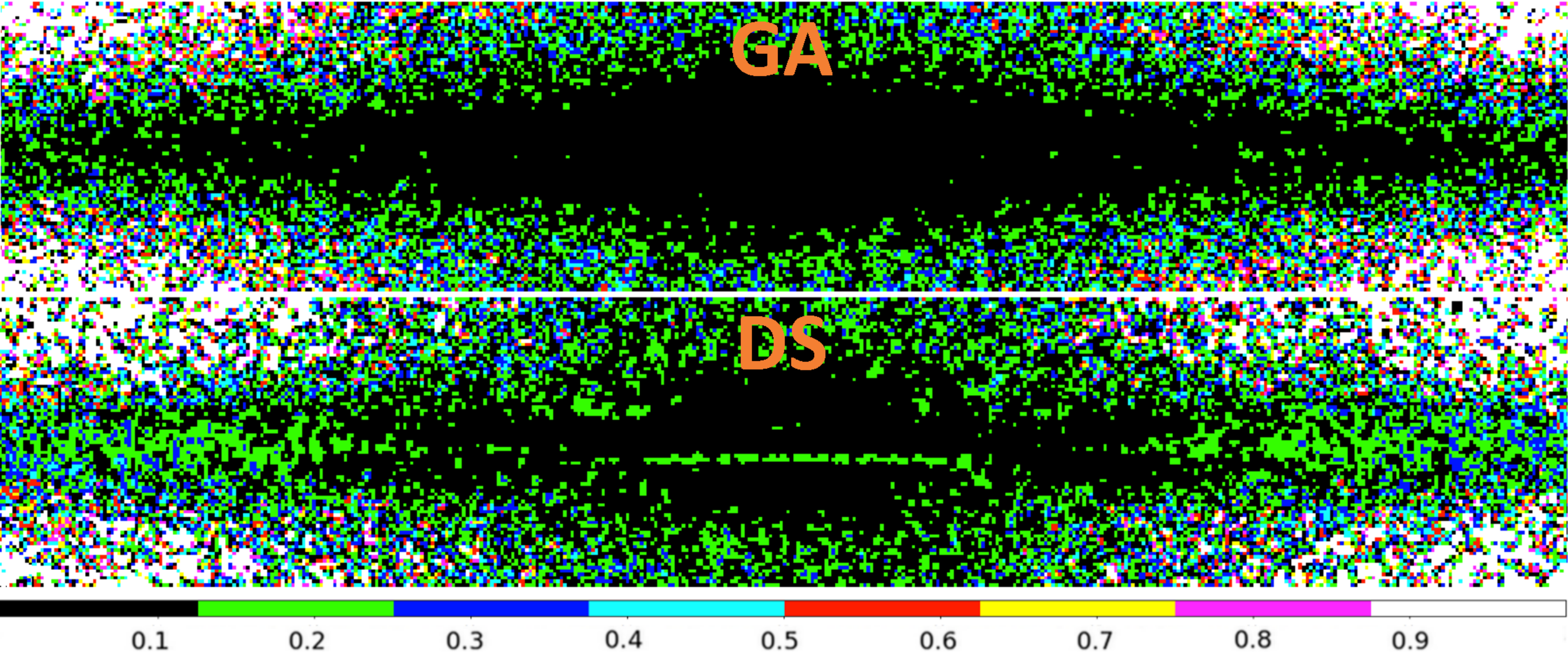}
 \caption{The residual frame of the reference image and the best fitting 
 model as determined by the GA (top) and the repeatedly restarted Downhill Simplex method (Bottom). The parameter values are listed in Table~{\ref{GA_DS.tab}}. The green line in the bottom image is the result of slight 
 inaccuracies in the reproduction of the dust lane.}
 \label{GA_DS_RES.pdf}
\end{figure*} 

As described in Section~{\ref{GAlib.sec}}, the choice of genetic
algorithms as our preferred optimization routine was driven by the
nature of our problem: the minimization of a complex, numerical and
noisy objective function in a large, multi-dimensional parameters
space, characterized by several local minima. As we have demonstrated,
the genetic algorithm approach turned out to be a powerful tool in
this respect, and enabled us to reach our goals. In particular,
concerning the problem of noisy objective functions, genetic
algorithms are very effective: since they work on a population rather
than iteratively on one point, that the random noise in the objective
function is a much less determining factor in the final solution
\citep{2000ApJ...545..974M, 2003AJ....125.1958L}. On the other hand,
the method is not entirely free of drawbacks: compared to some other
optimization routines, genetic algorithms have a relatively slow
convergence rate on simple problems, and the error analysis is not
easily treated in a natural way.

To further explore other possibilities and not limit ourselves to this
one approach only, we have performed further tests adopting two other
optimization schemes that have been used in radiative transfer fitting
problems, namely the Levenberg-Marquardt and the downhill-simplex
methods. Both methods were applied to the test case we have considered
in Section~{\ref{ElevenFree.sec}}, and their performance is compared
to genetic algorithms. In the following we briefly describe these two
routines and present the results of the comparison.

\begin{table*}
  \centering
  \begin{tabular}{|c|c|c|c|c|}
    \hline
    & & & & \\
    parameter & unit & reference  & GA & DS\\ 
    & & & & \\ \hline 
    & & & & \\
    $M_{\text{d}}$ & $10^7~M_\odot$ & 4 & $3.75\pm0.5$ & $4.57 \pm1.61$ \\
    $\tau_{\text{f}}$ & -- & 0.80 & $0.98 \pm 0.14$ & $1.26 \pm 0.37$\\ 
    $h_{R,{\text{d}}}$ & kpc & 6.6 & $5.83\pm0.75$ &$7.90 \pm 1.44$\\
    $h_{z,{\text{d}}}$ & pc & 250 & $245\pm30$ &$284 \pm 61$\\
    $L_{\text{tot}}$ &$10^{9}\ L_{\odot}$ & 1 & $0.97 \pm 0.09$ &$0.98 \pm 0.13$\\
    $h_{R,*}$ & kpc & 4.4 & $4.36\pm0.26$ &$4.0\pm 0.5$\\
    $h_{z,*}$ & pc & 500 & $519\pm51$ &$505 \pm 84$\\
    $B/D$ & -- & 0.33 & $0.34\pm0.08$ &$0.23 \pm 0.19$\\ 
    $R_{\text{e}}$ & kpc & 2 & $1.75\pm0.24$ &$2.16\pm0.45$\\
    $n$ & -- & 2.5 & $2.5\pm0.5$ &$3.52 \pm 0.65$\\ 
    $q$ & -- & 0.5 & $0.50\pm0.04$ &$0.51 \pm 0.17$\\
    $i$ & deg & 89 & $88.9\pm0.1$ &$89.2 \pm 0.3$\\ 
    & & & & \\
    \hline 
  \end{tabular}
  \vspace*{2ex}
  \caption{Input values and fitted values of the parameters of the
    model used in the test simulations in Section~{\ref{TestImages.sec}}. The values used
    to create the reference image (see Figure~{\ref{TestImage.pdf}})
    are given in the third column. The fourth an fifth column
    list the fitted values for these
    parameters, together with their 1$\sigma$ error bars, for the fits
    with eleven free parameters for the genetic algorithms and downhill simplex method respectively.}
  \label{GA_DS.tab}
\end{table*}

\subsection{Levenberg-Marquardt method}

The Levenberg-Marquardt algorithm or damped least-squares method
\citep{Levenberg44, Marquardt63} is an iterative optimization method,
that uses the gradient of the objective function and a damping factor
based on the local curvature to find the next step in the iteration
towards the final solution. The Levenberg-Marquardt method is one of
the most widely used methods for nonlinear optimization problems. It
has been applied in inverse radiative transfer modelling by
\citet{1999A&A...344..868X} and \citet{2007A&A...471..765B}. The
choice of this algorithm was possible because the radiative transfer
approach used by both was an (approximate) analytical model
(\citet{2007A&A...471..765B} only used the Levenberg-Marquardt method
for fits with ray-tracing radiative transfer simulations where
scattering was taken into account). In that case, the gradient in the
multi-dimensional parameter space only requires a modest computational
effort and is free of random noise.

As we are dealing with a Monte Carlo radiative transfer code with a
full inclusion of multiple anisotropic scattering, adapting this
method was not trivial. In particular, the presence of Monte Carlo
noise on the objective function evaluation makes the (numerical)
calculation of the gradient difficult. In an attempt to overcome this
problem we have combined two solutions:
\begin{itemize}
\item we increase the number of photon packages in every Monte Carlo
  simulation, so that the noise is reduced to a minimum but leads to
  computationally more demanding simulations;
\item we calculate the gradient based on different different objective
  function evaluations in every parameter space dimension. This helps
  to mitigate both the high impact that the noise has on the gradient
  in regions close to the starting point, and also the unrealistic
  values that the gradient can yield when calculated on points located
  further away from the current position.
\end{itemize}
In our implementation, we found that the number objective function
evaluations needed to make one single Levenberg-Marquardt step is
around 70 for our optimization problem containing 11 free parameters
(Section~{\ref{ElevenFree.sec}}). Keeping in mind that we have to use
a larger number of photon packages, one single Levenberg-Marquardt
step is comparable to approximately 400 evaluations performed in the
genetic algorithms approach, in terms of computational effort.

\subsection{Downhill simplex method}
 
The downhill simplex method, also known as Nelder-Mead method or
amoeba method \citep{NelderMead65}, is another commonly used
non-linear optimization technique. The method iteratively refines the
search space defined by an ($N+1$)-dimensional simplex by replacing
one of its defining points, typically the one with the worst objective
function evaluation, with its reflection through the centroid of the
remaining $N$ points defining the simplex.

In relatively simple optimization problems, the downhill simplex
method is known to have a relatively poor convergence, i.e.\ it is not
efficient in the number of objective function evaluations necessary to
find the extremum. On the positive side, the method only needs the
evaluation of the objective function itself and does not require
additional information on the parameter space or the calculation of
gradients.  Importantly, the method also works better with noisy
objective functions compared to Levenberg-Marquardt. Hence, coupling a
downhill simplex routing to a Monte Carlo radiative transfer code
requires no additional adjustments as instead needed in the case of
Levenberg-Marquardt. The downhill simplex method was used by
\citet{2007A&A...471..765B} for the general case, i.e.\ for fits in
which scattering was included (and hence Monte Carlo noise was
present).

\subsection{Results and comparison}

When applying the Levenberg-Marquardt and downhill simplex algorithms
to the problem discussed in Section~{\ref{ElevenFree.sec}}, both
methods are initialized in the same, randomly determined position and
new evaluations are performed until there is no significant
improvement in the values of the objective function. The left panel of
Figure~{\ref{D_S2.pdf}} shows the function evaluations (points) and
the averaged path (lines) of the Levenberg-Marquardt (green) and
downhill simpex (blue) methods for the dust mass and scale length of
the dust disk, two parameters which are hard to constrain separately.
The first evaluation --or starting point-- is marked with a black open
circle, while the ``real'' values are indicated by the red lines. The
history of the function evaluations is colour coded (from black at the
beginning to gradually green or blue, for the last evaluation).

Figure~{\ref{D_S2.pdf}} visually demonstrates another fundamental
complication of both the Levenberg-Marquardt and downhill simplex
methods: they both fail to reach the area of the parameter space
located closely to the true values. Even for parameters which are more
easily determined, like the scale length and height of the stellar
disk (see the right panel in Figure~{\ref{D_S2.pdf} }), the methods do
not succeed at reaching the region around the true values. This,
however, should not come as a surprise since both methods are known to
be local rather than global optimization methods. This limitation can
be partially overcome by repeatedly restarting the algorithm in a new
random initial position. For the Levenberg-Marquardt method, this is
simply unfeasible as it would result in a computationally exhausting
task, as the number of photon packages used in every Monte Carlo
simulation had to be increased substantially to mitigate the effects
of Monte Carlo noise. It was therefore not considered to be a viable
method for this problem.

To compare the downhill simplex method to genetic algorithms for our
specific problem, we have followed the afore mentioned idea,
repeatedly restarting the downhill simplex optimization at random
positions within the same boundaries, until the same number of
function evaluations was reached. Table~{\ref{GA_DS.tab}} lists the
values of the best fit model, together with the error determined by
the standard deviation of the five best models. In
Figure~{\ref{GA_DS_RES.pdf}} we show the residuals of the best fit models
determined by the downhill simplex and genetic algorithms
approaches. Comparing both the residuals and the best-fit parameter
values, it is clear the downhill simplex method is not capable of
reproducing the nominal values with the same degree of accuracy
reached by the genetic algorithm approach.  Also, the spread in the
parameters of the final solutions is considerably larger compared to
the values obtained using the genetic algorithms. Again, this is the
result of the ``local'' nature of the downhill simplex method. Genetic
algorithms generally perform much better at both determining the
global minimum and at handling the noise, turning out to be the most
efficient and robust choice to solve this problem.

\end{appendix}
}

\end{document}